\documentclass[12pt]{article}
\usepackage{amsfonts,amssymb,cite}

\newcommand{\be}{\begin{equation}}
\newcommand{\ee}{\end{equation}}
\newcommand{\bea}{\begin{eqnarray}}
\newcommand{\eea}{\end{eqnarray}}
\newcommand{\beas}{\begin{eqnarray*}}
\newcommand{\eeas}{\end{eqnarray*}}

\begin{document}

\baselineskip 14 pt
\parskip 12 pt

\begin{titlepage}
\begin{flushright}
{\small BROWN-HET-1359} \\
{\small CU-TP-1088} \\
{\small hep-th/0306209}
\end{flushright}

\begin{center}

\vspace{2mm}

{\Large \bf Stretched horizons, quasiparticles and quasinormal modes}

\vspace{3mm}

Norihiro Iizuka${}^1$, Daniel Kabat${}^1$, Gilad Lifschytz${}^2$ and David A.\ Lowe${}^3$

\vspace{2mm}

${}^1${\small \sl Department of Physics} \\
{\small \sl Columbia University, New York, NY 10027} \\
{\small \tt iizuka, kabat@phys.columbia.edu}
\vspace{1mm}

${}^2${\small \sl Department of Mathematics and Physics and CCMSC} \\
{\small \sl University of Haifa at Oranim, Tivon 36006, ISRAEL} \\
{\small \tt giladl@research.haifa.ac.il}
\vspace{1mm}

${}^3${\small \sl Department of Physics} \\
{\small \sl Brown University, Providence, RI 02912} \\
{\small \tt lowe@brown.edu}

\end{center}

\vskip 0.3 cm

\noindent
We propose that stretched horizons can be described in terms of a gas
of non-interacting quasiparticles.  The quasiparticles are unstable,
with a lifetime set by the imaginary part of the lowest quasinormal
mode frequency.  If the horizon arises from an AdS/CFT style duality
the quasiparticles are also the effective low-energy degrees of
freedom of the finite-temperature CFT.  We analyze a large class of
models including Schwarzschild black holes, non-extremal D$p$-branes,
the rotating BTZ black hole and de Sitter space, and we comment on
degenerate horizons.  The quasiparticle description makes manifest the
relationship between entropy and area.

\end{titlepage}

\section{Introduction}

The physics of black holes has attracted wide attention as an arena
for testing theories of quantum gravity.  In particular a great deal
of effort has gone towards resolving the puzzles of black hole
thermodynamics and information loss.  In this context the idea of a
stretched horizon arose as a useful tool for thinking about black
holes.  It originated as a classical description of black holes as
seen by outside observers \cite{MembraneParadigm}, but the concept was
later borrowed to help give a consistent quantum mechanical
interpretation of black hole physics \cite{stu}.

Since we now have a complete non-perturbative description of certain
spacetimes in the form of the AdS/CFT correspondence and its
generalizations \cite{AdSCFTreview} it is natural to see if the idea
of a stretched horizon is correct. This is not just a hypothetical
question, since if one understands how to use the stretched horizon
idea, one will gain insight into both the structure of black holes and
the structure of the dual quantum field theory.

Recently it was shown that a mean field approximation to the strongly
coupled gauge theory which describes non-extremal 0-brane black holes
gives results consistent with supergravity expectations
\cite{KLL1,KLL2,Iizuka:2001cw}. This led to the idea that these
finite-temperature gauge theories are well described by a gas of
non-interacting quasiparticles with energies of order the temperature
and lifetimes of order the inverse temperature.  It was then shown
that a stretched horizon description (a purely space time perspective)
has exactly the same properties \cite{quasi_I}.  It was also shown
that the stretched horizon description correctly accounts for the rate
at which energy is thermalized in the gauge theory.  This agreement
relies on putting the stretched horizon one Planck proper distance
away from the event horizon.  One goal of the present paper is to
understand this better, since the classical notion of a stretched
horizon does not fix where it resides \cite{MembraneParadigm}, while
quantum mechanically it has been placed where the proper temperature
matches either the Planck \cite{stu} or string \cite{Sen:1995in}
scale.

The theme of this paper is to show that the degrees of freedom of a
stretched horizon can be viewed as a gas of quasiparticles.  The
quasiparticle picture has the advantage of being extremely simple.  It
also makes manifest several universal properties of horizons,
including the universal relationship between entropy and horizon area.

In developing this picture of a stretched horizon we have drawn
heavily from several sources.  The basic idea of a stretched horizon
originated with the membrane paradigm \cite{MembraneParadigm}.
Stretched horizons were used to describe the quantum properties of a
black hole by Susskind, Thorlacius and Uglum \cite{stu}, who pointed
out several universal thermodynamic properties of the stretched
horizon which are fundamental to the quasiparticle description.
Stretched horizons were also used by Sen \cite{Sen:1995in} to count
the states of a family of degenerate black holes.  Although we will
not be able to make a precise connection, we find it inspirational to
identify the quasiparticles with the open strings stuck on the horizon
introduced by Susskind \cite{Susskind}.  Finally, we note that the
quasiparticle description we will present has some overlap with the
work of York \cite{york}.

An outline of this paper is as follows.  We first show how the
quasiparticle picture works for Schwarzschild black holes.  We then
study non-extremal D$p$-branes, as a straightforward generalization of
our previous work on D0-branes \cite{quasi_I}.  In this context the
quasiparticles give information about the dual $p+1$ dimensional gauge
theory.  We then analyze a model with more parameters, namely the
rotating BTZ black hole \cite{BTZ}, and show how the stretched horizon
picture relates to the dual CFT.  The BTZ analysis should serve as a
good guide for working out quasiparticle properties in more
complicated situations.  Finally we show that the static patch of de
Sitter space can be described by a stretched horizon with sensible
quasiparticle properties.  We also comment on the location of the
stretched horizon and explain that for degenerate horizons it does not
need to be located at the Planck scale.  In an appendix we discuss
decoupling limits, quasinormal modes and the quasiparticle description
for a general class of $p$-brane solitons \cite{CveticYoum,lupope,kt}.

\section{Schwarzschild black holes}
\label{SchwarzschildSection}

As our first example, we show that a simple quasiparticle picture can
capture the basic properties of a Schwarzschild black hole.

In $d$ spacetime dimensions the Schwarzschild metric reads
\[
ds^2 = - f(r) dt^2 + {1 \over f(r)} dr^2 + r^2 d\Omega_{d-2}^2
\]
where
\[
f(r) = 1 - {\omega_d M \over r^{d-3}} \qquad \omega_d = {16 \pi G \over (d-2) {\rm vol}(S^{d-2})}\,.
\]
The parameter $M$ is the mass of the black hole.  The event horizon is
located at the Schwarzschild radius $R_S = (\omega_d M)^{1/(d-3)} \sim
(GM)^{1/(d-3)}$.  The Hawking temperature is
\[
T = {d - 3 \over 4 \pi R_S}
\]
and the Bekenstein-Hawking entropy is
\[
S = {A \over 4 G} = {{\rm vol}(S^{d-2}) R_S^{d-2} \over 4 G} \sim (\ell_{\rm Planck} M)^{(d-2)/(d-3)}\,.
\]

We will introduce a stretched horizon \cite{MembraneParadigm} at a
radius where the proper temperature $T_{\rm proper} = T /
\sqrt{-g_{tt}}$ is equal to the Planck temperature.  As we discuss
below, this is necessary to get a consistent description.  The
Stefan-Boltzmann law gives the rate at which the stretched horizon
emits energy in outgoing Hawking radiation.\footnote{In equilibrium
this emission rate is balanced by an equal and opposite flux of
infalling radiation.  Also note that only a tiny fraction of this
emitted radiation ever reaches asymptotic infinity; the vast majority
of Hawking particles turn around and fall back onto the stretched
horizon.  See p.~287 of \cite{MembraneParadigm}.}
\be
\label{SB1}
{d E_{\rm proper} \over dt_{\rm proper}} \sim A T_{\rm proper}^d
\ee
Multiplying this equation by $-g_{tt}$ gives the emission rate
measured with respect to the Schwarzschild time coordinate $t$.
The proper temperature $\sim 1/\ell_{\rm Planck}$, so
\be
\label{SchwarzschildRate}
{d E \over dt} \sim {A \over \ell_{\rm Planck}^{d-2}} \, T^2\,.
\ee

We now postulate that the stretched horizon of the black hole can be
approximately described as a gas of $N$ non-interacting quantum
mechanical degrees of freedom (``quasiparticles'').  The number of
quasiparticles depends on the temperature, as we postulate that the
total number of quasiparticles matches the entropy of the black hole,
$N \sim S$.  Each quasiparticle has an energy $\epsilon \sim T$ and
decays with a characteristic lifetime $\tau \sim 1/T$.  We take the
quasiparticle gas to be in thermal equilibrium at temperature $T$.

It is easy to see that the quasiparticle gas reproduces the
equilibrium thermodynamic properties of the black hole.  The energy of
the gas
\[
E \approx N \epsilon \sim M
\]
agrees with the mass of the black hole, and by construction the
entropy of the gas is equal to the Bekenstein-Hawking entropy.  These
properties of the quasiparticle gas correspond to the universal
properties of the stretched horizon noted by Susskind, Thorlacius and
Uglum \cite{stu}.  But the quasiparticle gas also reproduces certain
dynamical properties of the black hole, since the rate at which energy
gets redistributed in the gas due to the ongoing production and decay
of quasiparticles is given by
\be
\label{QuasiRate}
{d E \over dt} \approx N \epsilon / \tau \sim N T^2\,.
\ee
This agrees with the rate (\ref{SchwarzschildRate}) at which energy gets
redistributed on the stretched horizon due to the emission and absorption
of Hawking quanta.

Let us make a few comments on these results.  Quite generally we
expect that any horizon can be described in terms of a membrane
located at the stretched horizon.  This fact is well-established at
the classical level \cite{MembraneParadigm}, and has been argued to
hold in the quantum theory as well \cite{stu}.  We are claiming that
in the quantum theory the membrane has a low-energy effective
description in terms of a gas of quasiparticles.  Although we cannot
make any sort of precise connection, we would like to identify the
quasiparticles with the open strings stuck on the horizon introduced
by Susskind \cite{Susskind}.

We expect the quasiparticle gas to provide a holographic description
of the black hole together with certain of its low-energy excitations.
If the black hole spacetime admits a dual description along the lines
of the AdS/CFT correspondence \cite{AdSCFTreview} then one should also
be able to identify the quasiparticles in the dual field theory.  In
\cite{quasi_I} we began with a dual field theory of 0-branes and used
mean-field methods to extract an effective quasiparticle description.
The resulting quasiparticle gas indeed captures the thermodynamic properties
of the black hole \cite{KLL1,KLL2,Iizuka:2001cw}.

In the case of a Schwarzschild black hole no dual description is
known,\footnote{See however \cite{Danielsson}.  Also note that
Schwarzschild black holes are present as excited states in the AdS/CFT
correspondence; perhaps the quasiparticles provide an effective
description of a sector of the CFT Hilbert space. See \cite{lowebh}
for some steps in this direction.} so we are simply
postulating that the quasiparticle gas has the right properties to
match the black hole.  Some peculiar features of the Schwarzschild
black hole, for example its negative specific heat, are simply
reflected in the fact that the number of quasiparticles goes down as
the temperature increases.

The quasiparticle description does give insight into one key aspect of
black hole physics.  Note that the black hole emits energy at a rate
(\ref{SchwarzschildRate}) which is proportional to the area of the
event horizon in Planck units.  The quasiparticle gas, on the other
hand, redistributes energy at a rate which is proportional to the
number of quasiparticles.  The rates agree provided the entropy of the
quasiparticle gas is equal to the area of the horizon in Planck units.
This provides a simple origin for the entropy -- area relation, which
is otherwise very obscure from the point of view of any dual field
theory description \cite{quasi_I}.

We conclude our treatment of Schwarzschild black holes with a
discussion of the way in which a perturbed black hole returns to
(approximate) thermal equilibrium.\footnote{The black hole is of
course ultimately unstable due to evaporation.  The quasiparticle
description is only valid, and the notion of equilibrium only applies,
on timescales short compared to the black hole lifetime.}  The
quasiparticle gas should describe low-lying excitations of the
stretched horizon.  From the quasiparticle point of view such
perturbations should decay on a timescale set by the quasiparticle
lifetime $\tau \approx 1/T$.  On the other hand, from the gravity
point of view such perturbations can be described as quasinormal
excitations of the black hole \cite{QuasinormalReview}.  Quasinormal
excitations of Schwarzschild black holes have been studied recently in
\cite{Motl1,CardosoSchwarzschild,Motl2,Birmingham}.  The decay of scalar perturbations, for
example, is governed by the wave equation
\beas
& & \left(- f(\rho) \partial_\rho f(\rho) \partial_\rho + V(\rho) - (R_S \omega)^2\right) \psi = 0 \\
& & f(\rho) = 1 - 1 / \rho^2 \\
& & V(\rho) = {f \over \rho^2} \left[\ell(\ell + d - 3) + {(d-2)(d-4) \over 4} + {(d-2)^2 \over 4 \rho^{d-3}}\right]
\eeas
where $\rho = r / R_S$ is dimensionless.  The quasinormal frequencies
$\omega$ are determined by solving this equation with boundary
conditions that are pure ingoing at the horizon and pure outgoing at
infinity.  It is obvious on dimensional grounds that
\[
\omega \sim 1/R_S \sim T\,.
\]
The same dimensional analysis works for vector and gravitational
perturbations of the black hole \cite{Birmingham}.

Note that these quasinormal modes are supergravity solutions which
govern the late-time behavior of certain bulk closed-string fields
\cite{QuasinormalReview}.  They should not be confused with the
quasiparticles, which we regard as arising from open string degrees of
freedom on the stretched horizon.  But we expect the quasiparticles to
provide a holographic description of low-energy excitations near the
black hole.  In particular the quasiparticle lifetime should govern
the decay of the low-lying quasinormal modes.  Thus we propose that
the quasiparticle inverse lifetime is of order the imaginary part of
the lowest quasinormal frequency, $\tau \sim 1/T$.\footnote{See
\cite{york} for an earlier investigation along these lines. It would
be interesting to relate this identification to the proposal discussed
in \cite{hod,dreyer,Motl1}.}

As in \cite{quasi_I}, we can use this result for the quasinormal
frequencies to show that the stretched horizon must be placed at the
Planck temperature.  The Stefan-Boltzmann law (\ref{SB1}) can be
turned into an expression for the rate at which the stretched horizon
emits entropy.
\[
{dS \over dt} \sim T \cdot A T_{\rm proper}^{d-2}
\]
Given the quasinormal lifetime we expect $dS/dt \sim T S$.  Thus we
must identify $A T_{\rm proper}^{d-2}$ with the black hole entropy,
which is only consistent provided we place the stretched horizon at
the Planck temperature.

\section{Non-extremal D$p$-branes}
\label{bbranes}

As our second example we discuss non-extremal black D$p$-branes in
type II string theory.  We take a near-horizon limit, so that the dual
field theory is known \cite{imsy}.  This section generalizes our
previous work on D0-brane black holes \cite{quasi_I}.

\subsection{Gravitational description}

We begin with the near-horizon limit of the supergravity solution
describing $N$ coincident non-extremal D$p$-branes \cite{imsy}.  The
string-frame metric and dilaton are
\beas
{1 \over \alpha'} ds^{2} & = & \frac{U^{(7-p)/2}}{\sqrt{d_{p}e^2}}\left(-(1-\frac{U_{0}^{7-p}}{U^{7-p}})dt^2
+dx_\parallel^{2}\right)+\frac{\sqrt{d_{p}e^2}}{U^{(7-p)/2}}(1-\frac{U_{0}^{7-p}}{U^{7-p}})^{-1}dU^2 \\
& & +\sqrt{d_{p}e^2} U^{(p-3)/2} d\Omega^{2}_{8-p} \\
e^\phi & = & (2\pi)^{(2-p)}g^2_{YM}\left(\frac{U^{7-p}}{d_{p}e^2}\right)^{(p-3)/4}
\eeas
where $e^2 = g_{YM}^2 N$ and $d_p = 2^{7-2p} \pi^{\frac{9-3p}{2}} \Gamma(\frac{7-p}{2})$.
The Einstein metric is
\bea
\nonumber
ds^{2} & = & {\rm const.} \, U^{(7-p)^{2}/8}\Biggl[-\left(1-\frac{U_{0}^{7-p}}{U^{7-p}}\right)dt^2
+dx_\parallel^{2} + \frac{d_{p}e^2}{ U^{7-p} (1-\frac{U_{0}^{7-p}}{U^{7-p}}) } dU^{2} \\
\label{Einstein}
& & \quad +d_{p}e^2 U^{p-5} d\Omega^{2}_{8-p}\Biggr]\,.
\eea
Up to inessential numerical factors the Hawking temperature, energy
density and entropy density depend on the horizon radius $U_0$ according to
\bea
\label{p-braneTemp}
T & \sim & {1 \over e} U_0^{(5-p)/2} \\
\nonumber
E & \sim & {N^2 \over e^4} U_0^{7-p} \\
\nonumber
s & \sim & {N^2 \over e^3} U_0^{(9-p)/2}
\eea

As in the previous section, we introduce a stretched horizon at a
radius where the proper temperature has the value $1/\ell_{\rm
Planck}$.  The proper rate at which the stretched horizon radiates
energy density is given by Stefan-Boltzmann,
\[
{d E_{\rm proper} \over dt_{\rm proper}} = A T_{\rm proper}^{10-p}
\]
where $A$ is the proper area of an $(8-p)$-sphere with radius $U_0$.  Converting
to Schwarzschild quantities this means
\be
\label{p-braneRadiation}
{d E \over dt} = {A \over \ell_{\rm Planck}^{8-p}} T^2\,.
\ee

\subsection{Quasinormal modes}
\label{p-braneQuasi}

We now study quasinormal excitations of the black brane metric
(\ref{Einstein}).  Our approach will parallel the study of AdS
Schwarzschild black holes in \cite{hubeny, CardosoAdS}.  For
simplicity we study a massless minimally-coupled scalar field.
Separating variables
\be
\phi=e^{-i\omega t+ik_{i}x^{i}} f(U) Y(\Omega)
\ee
where $Y(\Omega)$ is an eigenfunction of the Laplacian on $S^{8-p}$ with
eigenvalue $-l(l+7-p)$, the scalar wave equation becomes
\beas
&&\partial_{U}U^{8-p}\left(1-\frac{U_{0}^{7-p}}{U^{7-p}}\right)\partial_{U}f(U)+
d_{p}e^2 U\left(\frac{\omega^2}{1-\frac{U_{0}^{7-p}}{U^{7-p}}}-k^2\right)f(U) \\
&&\qquad-l(l+7-p) U^{6-p}f(U)=0~.
\eeas
We now change variables to
\be
z=1-\frac{U_{0}^{7-p}}{U^{7-p}}
\ee
which gives the equation
\be
\left(z\partial_{z}z\partial_{z}+\frac{\rho^{2}-z\kappa^{2}}{(1-z)^{\frac{9-p}{7-p}}}-\frac{l(l+7-p)z}{(1-z)^{2}(7-p)^{2}}\right)
f(z)=0
\label{numer}
\ee
where 
\be
\rho^{2}=\frac{\omega^{2}d_{p}e^2}{(7-p)^{2}U_{0}^{5-p}},\ \ 
\kappa^{2}=\frac{k^{2}d_{p}e^2}{(7-p)^{2}U_{0}^{5-p}}~.
\ee
We solve (\ref{numer}) with the Dirichlet boundary condition \cite{hubeny}
\be
\label{rooteq}
f(z=1)=0
\ee
at infinity and the purely ingoing condition at the future horizon
\be 
\label{horbc}
f(z\to 0) \sim z^{-i \rho}~.
\ee
For given values of $l$ and $\kappa$ there are solutions only for
discrete values of $\rho$, labelled by an integer $n$.
The $n^{th}$ quasinormal frequency is then
\be
\omega_n = \frac{(7-p)U_{0}^{\frac{5-p}{2}}}{\sqrt{d_{p}e^2}} \rho_n\,.
\ee
Note that the quasinormal frequencies are proportional to the Hawking
temperature, since
\be
\omega_n \sim {1 \over e} U_0^{(5-p)/2} \sim T\,.
\ee
To obtain the coefficient of proportionality we must solve
(\ref{numer}) numerically.  We focus on the mode with lowest imaginary
part, corresponding to $\kappa=0$ and $l=0$. For $p<5$, we use a power
series expansion around $z=0$ satisfying the boundary condition
(\ref{horbc}) matched at a regular intermediate point (e.g. $z=1/2$)
with a Runge-Kutta solution to (\ref{numer}) evolved from $z=1$, and
satisfying (\ref{rooteq}).  We solve for $\rho$ by demanding that
$f'(z)/f(z)$ coincide at the matching point. This method is
implemented using Mathematica as described in more detail in appendix
A.  This method yields dramatically faster convergence than the pure
power series solution of \cite{hubeny}, producing the results quoted
here in just a few seconds. The difficulty with the pure power series
solution is that the function is evaluated at its radius of
convergence ($z=1$), so convergence of the series can be very slow.
The results for the lowest modes are shown in the following table.
\begin{equation}
\begin{array}{lc}
p & \rho \cr
\noalign{\vskip 2mm}
0\ & 0.554059-0.930000i \cr
1\ & 0.616474-0.887951i \cr
2\ & 0.693098-0.816466i \cr
3\ & 0.779863-0.686669i \cr
4\ & 0.834579-0.435416i \cr
\end{array}
\end{equation}
For $p=1,\,3,\,4$ these frequencies agree with the results of
\cite{hubeny} for large AdS-Schwarzschild black holes in
$AdS_{4,\,5,\,7}$ respectively.\footnote{To see this note that the
wave equation (\ref{GeneralWave}) in appendix B is invariant under
shifting $D \rightarrow D+1$ and $p \rightarrow p+1$, and that in
these cases the metric (\ref{newmet1}) agrees with (2.8) in
\cite{hubeny} upon setting $\rho = 1/r$.}

For $p=5$, (\ref{numer}) can be solved analytically in terms of
hypergeometric functions. The purely ingoing (for $Im(\rho) <0$)
solution on the horizon takes the form
\be
\label{hyperg}
z^{-i \rho} (1-z)^{\gamma}{}_2 F_1 (\gamma-i \rho, \gamma - i \rho,
1-2 i \rho; z)
\ee
where we have defined
\be
\gamma = \frac{1+\sqrt{1-4 (\rho^2 -\tilde{\kappa}^2)}}{2} ~, \qquad
\tilde \kappa ^2 = \kappa^2 + \frac{l(l+2)}{4}~.
\ee
Near infinity $z\to 1$, the solution (\ref{hyperg}) is a linear
combination of a component that is asymptotically $(1-z)^\gamma$, and
one that is $(1-z)^{1-\gamma}$. To get a quasinormal mode with finite
flux at infinity, we need to set the coefficient of one of these terms
to zero. Therefore we must solve either
\be
\label{terma}
\frac{ \Gamma(1-2 i \rho)\Gamma( 2\gamma -1) }{ \Gamma(
\gamma-i \rho)^2} =0
\ee
or
\be
\label{termb}
\frac{ \Gamma(1-2 i \rho)\Gamma(1-2\gamma) }{ \Gamma(1
-i\rho - \gamma)^2} =0~.
\ee
Equation (\ref{terma}) is satisfied when $ \gamma-i\rho = -n$,
with $n$ a positive integer, however this equation has no
solutions. Equation (\ref{termb}) has solutions when $1-i \rho -
\gamma = -n$ with $n$ a positive integer. This has purely imaginary
solutions for $\rho$, but these always lead to divergent flux at
infinity.  However if $\rho$ is pure imaginary, then the flux on the
horizon vanishes, so one could imagine including an admixture of
\be
\label{hypergb}
z^{i \rho} (1-z)^{\gamma}{}_2 F_1 (\gamma+i \rho, \gamma +i \rho,
1+2 i \rho; z)~.
\ee
We still demand the solution be a function only of $e^{-i \omega t}
z^{-i \rho}$ as $z\to 0$, which rules out this component.  We conclude
that there are no well-behaved quasinormal modes for black 5-branes.

For $p=6$ the equation does not appear to be soluble analytically.
The solutions have oscillatory behavior as $z\to 1$, so rather
different numerical methods are needed to obtain quasinormal mode
frequencies reliably. The methods used for Schwarzschild black holes
should be useful, since when uplifted to eleven dimensions the $p=6$
solution is asymptotically flat \cite{imsy}. We leave further analysis
of this case for future work.

\subsection{Quasiparticle picture for $p \leq 4$}

For $p \leq 4$ it is easy to reproduce the thermodynamic properties of
the black brane by introducing a quasiparticle gas.  We postulate that
the longitudinal number density of quasiparticles is equal to the
entropy density of the black brane, $n \sim s$.  Each quasiparticle
has an energy $\epsilon$ of order the temperature and a lifetime
$\tau$ set by the imaginary part of the lowest quasinormal frequency.
Given our results in section \ref{p-braneQuasi}, this means $\tau \sim
1/T$.

It is easy to see that the energy density and entropy density of the
quasiparticle gas agree with the corresponding properties of the black
brane.  Moreover the rate at which the quasiparticle gas redistributes
energy density
\[
{d E \over dt} = n \epsilon / \tau
\]
agrees with the black brane radiation rate (\ref{p-braneRadiation}).
Note that for D$p$-branes the energy and entropy are power-law in the
temperature.  This property helps make a simple quasiparticle
description possible.

Finally, following \cite{quasi_I}, we show that the quasiparticle
picture correctly predicts the horizon radius of a black $p$-brane.
From the point of view of the dual $(p+1)$-dimensional gauge theory
one would define the radius of the black brane by
\[
R_h^2 = {1 \over N} \langle {\rm Tr} X^2 \rangle\,.
\]
Here $X$ is one of the transverse matrix-valued scalar fields, related
to a canonically normalized field by $X = e Y / \sqrt{N}$.  The
effective number of entries in the matrix $Y$ is given in terms of the
entropy density by $N_{\rm eff} = s / T^p$.  Thus
\[
R_h^2 = {e^2 \over N^2} N_{\rm eff} \langle y^2 \rangle
\]
where $\langle y^2 \rangle$ measures the fluctuation of a single
canonically normalized scalar field.  Upon suitably smearing the field
to suppress high-frequency contributions
\cite{Iizuka:2001cw,quasi_I,LennyFlat}, this correlator is given by
$\langle y^2 \rangle \sim T^{p-1}$.  Thus the quasiparticle picture
predicts
\[
R_h^2 \sim {e^2 s \over N^2 T}\,.
\]
This is indeed satisfied for black $p$-branes, upon identifying $R_h$
with $U_0$.

\subsection{Comments on $p=5$}

For D5-branes we were not able to find quasinormal modes, and a simple
quasiparticle description does not appear to be possible.  In this
case the temperature (\ref{p-braneTemp}) is independent of the energy
density at leading semiclassical order.  Sub-leading one-loop effects
drive the specific heat negative and indicate that the thermodynamic
ensemble is unstable \cite{Kuta}.  As described in \cite{Real}, a
classical instability is present for the D5-brane all the way down to
extremality, and this is argued to be equivalent to local
thermodynamic instability\footnote{It has been argued that there is a
correspondence between thermodynamic instability and classical
instability whenever a non-compact translational symmetry is present
\cite{Real}. Thus the Schwarzschild black hole is thermodynamically
unstable, but there is no translational symmetry, which is consistent
with the absence of a classical instability. As discussed in section
\ref{SchwarzschildSection}, the quasiparticle picture can make sense
here, because the timescale associated with thermalization ${\cal
O}(T^{-1})$ is much shorter than the timescale of evaporation ${\cal
O}(T^{1-d})$.  The $p<5$ D-branes are thermodynamically stable in the
decoupling limit, and as described in \cite{Real} have no classical
instability close to extremality (at least for the $p=1,2,4$ cases
studied there).}.  This has been conjectured to persist in the
decoupling limit \cite{Ranga}.  These facts are closely related to the
violation of the decoupling of the throat region and the 5-brane
described in \cite{mstrom}. The classical instability of the 5-brane
implies that thermalization will not occur, so the quasiparticle
picture will not work.

\section{BTZ black holes}
\label{BTZsection}

In this section we develop a quasiparticle description of the spinning
BTZ black hole \cite{BTZ}.  Unlike the previous cases, the BTZ black
hole depends on two parameters (mass and angular momentum) and has a
free energy which is not a simple power-law in the temperature.  This
will force us to introduce two distinct species of quasiparticles.

\subsection{Gravitational description}

Classical and quantum properties of the BTZ black hole are reviewed in \cite{Carlip}.
The metric is\footnote{We adopt units in which $8G = 1$.}
\be
\label{BTZmetric}
ds^2 = - V(r) dt^2 + {1 \over V(r)} dr^2 + r^2 \left(d\phi - {J \over 2 r^2}dt\right)^2
\ee
where $\phi$ has period $2\pi$ and
\[
V(r) = - M + {r^2 \over \ell^2} + {J^2 \over 4 r^2}\,.
\]
The mass $M$ and angular momentum $J$ are related to the horizon radii $r_\pm$ by
\beas
M & = & (r_+^2 + r_-^2)/\ell^2 \\
J & = & 2 r_+ r_- / \ell
\eeas
where $\ell$ is the asymptotic AdS radius.  The horizon rotates with angular velocity
\[
\Omega_H = J / 2 r_+^2 = r_- / \ell r_+\,.
\]
The Hawking temperature of the black hole is
\[
T = {r_+^2 - r_-^2 \over 2 \pi r_+ \ell^2}
\]
and the Bekenstein-Hawking entropy is
\[
S = 4 \pi r_+ \,.
\]
In the extremal limit a BTZ black hole has $M = J / \ell$ (or
equivalently $r_+ = r_-$) and vanishing temperature.  For simplicity
we will only consider non-extremal black holes, although we make a few
comments on the extremal limit in section \ref{AdditionalComments}.

We will place the stretched horizon at a radius where the proper
temperature $T_{\rm proper} = T/\sqrt{V(r)}$ is equal to the Planck
temperature.  We wish to compute the rate at which the stretched
horizon emits energy and angular momentum in outgoing Hawking
radiation.\footnote{In equilibrium these outgoing fluxes are of course
balanced by equal and opposite infalling fluxes.}  To do this we
introduce a family of ``Rindler observers'' who sit on the stretched
horizon and rotate with exactly the angular velocity of the event
horizon $\Omega_H$.  A typical Rindler observer has a worldline
\[
t(\tau) = \tau \qquad r(\tau) = {\rm const.} \qquad
\phi(\tau) = {\rm const.} + \Omega_H \tau\,.
\]
Such observers perceive themselves to be at rest with respect to a
thermal bath at temperature $T_{\rm proper}$.  This is easiest to see
in a Euclidean formulation: setting $\phi' = \phi - \Omega_H t$ and
Wick rotating $t = -i t_E$, $\Omega_H = i \Omega_{HE}$ one is forced
to identify
\[
(t_E,r,\phi') \sim (t_E+1/T,r,\phi')
\]
in order to get a non-singular Euclidean metric \cite{Carlip}.  The
identification is purely in the Euclidean time direction, so Rindler
observers who sit at fixed $\phi'$ are instantaneously at rest with
respect to the thermal bath.

We should point out that these Rindler observers differ from the
fiducial observers (or zero angular momentum observers) introduced in
\cite{MembraneParadigm}, who rotate with the local frame-dragging
angular velocity, and who see the event horizon as rotating beneath
them.  Also note that, due to the asymptotically AdS nature of space,
a Rindler observer at any radius outside the event horizon moves along
a timelike trajectory.  For example Rindler observers at spatial
infinity co-rotate with the CFT.

To relate local measurements made by Rindler observers to the
corresponding conserved Schwarzschild quantities, introduce
orthonormal basis vectors $\lbrace \bar{e}_0, \bar{e}_r,
\bar{e}_\phi\rbrace$ where
\[
\bar{e}_0 \approx {1 \over \sqrt{V(r)}}\left(\partial_t + \Omega_H \partial_\phi\right)
\]
is tangent to the Rindler worldline,
\[
\bar{e}_r = \sqrt{V(r)} \, \partial_r
\]
points in the radial direction, and
\[
\bar{e}_\phi \approx {1 \over r} \partial_\phi
\]
points in the $\phi$ direction.  (We have only written the leading
behavior as $r \rightarrow r_+$.)  On average, a Rindler observer sees
the outgoing Hawking quanta as having an energy-momentum vector of the
form
\[
\bar{p} = E_{\rm proper} \left(\bar{e}_0 + \bar{e}_r\right)\,.
\]
The corresponding outgoing conserved Schwarzschild energy and angular momentum
are obtained by taking inner products with the Killing vectors
$\partial_t$ and $\partial_\phi$.
\[
E_{\rm out} = - \langle \bar{p}, \partial_t \rangle \qquad
J_{\rm out} = \langle \bar{p}, \partial_\phi \rangle
\]
Up to corrections which are subleading as $r \rightarrow r_+$, these quantities are given by
\bea
\label{RindlerSchwarzschild}
&& E_{\rm out} - \Omega_H J_{\rm out} \approx \sqrt{V(r)} E_{\rm proper} \\
\nonumber
&& J_{\rm out} \approx {\ell r_+ r_- \over r_+^2 - r_-^2} \sqrt{V(r)} E_{\rm proper}\,.
\eea
Here we are assuming that the position of the stretched horizon
$r_{\rm stretch} = r_+ + \Delta r$ is such that $\Delta r \ll r_+ -
r_-$.  This assumption is valid since we're staying away from the
extremal limit $r_+ = r_-$.

A Rindler observer sees the stretched horizon of the black hole
radiate energy isotropically at a rate given by the Stefan-Boltzmann
law,
\[
{d E_{\rm proper} \over dt_{\rm proper}} \sim A T_{\rm proper}^3
\]
where $A = 2 \pi r_+$ is the area of the event horizon.  Since we
place the stretched horizon at a radius corresponding to the Planck
temperature, this means
\[
{d E_{\rm proper} \over dt_{\rm proper}} \sim {A \over \ell_{\rm Planck}} T_{\rm proper}^2\,.
\]
Using the dictionary (\ref{RindlerSchwarzschild}), the relations $dt_{\rm
proper} = \sqrt{V(r)} dt$ and $dE_{\rm proper} = dE/\sqrt{V(r)}$, and
recalling  $8G = 1$, the Stefan-Boltzmann
law implies that
\bea
\label{BTZevap}
&& {d \over dt}\left(E - \Omega_H J \right) \sim A T^2 \\
\nonumber
&& {d \over dt} J \sim {\ell r_+ r_- \over r_+^2 - r_-^2} A T^2\,.
\eea
Note that these equations are equivalent to
\bea
\label{Universal1}
&& {d \over dt} S \sim S T \\
\nonumber
&& {d \over dt} J \sim J T
\eea
where we have held $\Omega_H$ fixed and used $d E - \Omega_H dJ = T dS$ in the first
line.  Likewise the rate at which energy is radiated can be written as
\be
\label{Universal2}
{d \over dt} E \sim (E - F) T
\ee
where $F = E - TS - \Omega_H J$.  These relations suggest that all thermodynamic quantities relax
to equilibrium on a time-scale set by $1/T$.

\subsection{Quasiparticle description}
\label{BTZquasiparticle}

The BTZ metric (\ref{BTZmetric}) is dual to a conformal field theory
on the boundary \cite{BrownHenneaux, Strominger, AdSCFTreview}.  We
propose that this dual CFT has a low-energy effective description in
terms of a gas of weakly-interacting quasiparticles, which are
sufficient to describe the black hole and certain of its low-energy
excitations.  Thus the quasiparticle gas shares some properties with
the effective string discussed in \cite{DasMathur}.

To motivate our quasiparticle picture, we define dimensionless right-
and left- temperatures in the usual way 
\[
r_\pm = \pi \ell (T_R \pm T_L)
\]
Then the black hole energy, entropy, and angular momentum can be split
into right- and left- contributions according to
\bea
\nonumber
&& M = 2 \pi^2 (T_R^2 + T_L^2) \\
\label{BTZthermo}
&& S = 4 \pi^2 \ell (T_R + T_L) \\
\nonumber
&& J = 2 \pi^2 \ell (T_R^2 - T_L^2)
\eea

In order to account for both the black hole mass and angular momentum
we must introduce two distinct species of quasiparticles.  Our precise
proposal is that there are $N_R \approx 4 \pi^2 \ell T_R$ right-moving
quasiparticles, each with energy $\epsilon_R \approx T_R$ and lifetime
$\tau_R \approx 1/T_L$.  We likewise propose that there are $N_L
\approx 4 \pi^2 \ell T_L$ left-moving quasiparticles each with energy
$\epsilon_L \approx T_L$ and lifetime $\tau_L \approx 1/T_R$. 

A simple argument for the proposed quasiparticle lifetimes is that a
right-moving excitation can only thermalize by scattering off a
left-mover.  This suggests that the thermalization rate for
right-movers is proportional to the number of left-moving
quasiparticles, or equivalently $\tau_R \sim 1/N_L \sim 1/T_L$.  A
more detailed justification is given in section \ref{BTZquasinormal}.

In the quasiparticle picture it is straightforward to estimate the
energy, entropy, and angular momentum of the CFT.
\bea
\nonumber
E_{\rm CFT} & \approx & N_R \epsilon_R + N_L \epsilon_L \\
\label{CFTthermo}
S_{\rm CFT} & \approx & N_R + N_L \\
\nonumber
J_{\rm CFT} & \approx & N_R \epsilon_R - N_L \epsilon_L
\eea
It is also straightforward to estimate the rate at which energy and
angular momentum get redistributed among the CFT degrees of freedom,
due to the ongoing production and decay of quasiparticles.
\bea
\label{CFTevap1}
{dE_{\rm CFT} \over dt_{\rm CFT}} & \approx & N_R \epsilon_R / \tau_R + N_L \epsilon_L / \tau_L \\
\nonumber
{dJ_{\rm CFT} \over dt_{\rm CFT}} & \approx & N_R \epsilon_R / \tau_R - N_L \epsilon_L / \tau_L
\eea

We now show that these properties of the quasiparticle gas match the
corresponding properties of the black hole.  The induced metric on the
boundary is conformal to $ds^2_{\rm CFT} = - dt^2/\ell^2 + d\phi^2$,
so the dictionary is $t_{\rm CFT} = t/\ell$, $E_{\rm CFT} = \ell E$,
$J_{\rm CFT} = J$.  Thus the quasiparticle results (\ref{CFTthermo})
predict that
\bea
\nonumber
E & \approx & T_R^2 + T_L^2 \\
S & \approx & \ell(T_R + T_L) \\
\nonumber
J & \approx & \ell(T_R^2 - T_L^2)
\eea
in agreement with the black hole results (\ref{BTZthermo}). 
Likewise
the quasiparticles predict that
\bea
\nonumber
{d E\over dt} & \approx & (T_R + T_L) T_R T_L / \ell \\
\label{CFTevap2}
{d J\over dt}  & \approx & (T_R - T_L) T_R T_L
\eea
which are easily seen to be equivalent to the black hole expressions
(\ref{BTZevap}).\footnote{Note that these semiclassical expressions
for the Hawking flux of energy and angular momentum are only valid if
$N_L \gg 1$ and $N_R \gg 1$.}

Note that from the quasiparticle point of view $dE/dt$ is proportional
to the number of quasiparticles $N_R + N_L \sim T_R + T_L$, and is
therefore proportional to the entropy of the quasiparticle gas.  On
the other hand from the BTZ point of view $dE/dt$ follows from the
Stefan-Boltzmann law and is proportional to the area of the horizon in
Planck units.  Thus the quasiparticle description correctly accounts
for the relationship between entropy and area.

\subsection{Correlation times and quasinormal frequencies}
\label{BTZquasinormal}

In section \ref{BTZquasiparticle} we simply postulated that the lifetimes of
the left- and right-moving quasiparticles are given by
\[
\tau_L = 1/T_R \qquad \tau_R = 1/T_L\,.
\]
We now provide some justification for this, from both the supergravity
and CFT points of view.\footnote{A study of horizon degrees of freedom
for the BTZ black hole along the lines of \cite{hod,dreyer} was made
in \cite{BCC}.}  Our discussion in this section closely follows the
work of Birmingham, Sachs and Solodukhin \cite{BSS1,BSS2}.

Quasinormal modes for bulk fields of various spins were studied in
\cite{BSS1,CardosoBTZ}.  Working in terms of the CFT time coordinate
one set of solutions is
\be
\label{RightQuasiMode}
\Phi(t,r,\phi) = e^{-ik(t-\phi)} e^{-t/\tau_R} f(r) \qquad k \in {\mathbb Z}
\ee
where
\be
\label{RightQuasi}
\tau_R = 1/4 \pi T_L(h_L + n) \qquad n = 0,1,2,\ldots
\ee
Here $h_L$ is the conformal dimension of the corresponding operator in
the CFT.  These solutions describe right-moving excitations which
decay on a timescale $\tau_R$.  Likewise there are solutions
\be
\label{LeftQuasiMode}
\Phi(t,r,\phi) = e^{-ik(t+\phi)} e^{-t/\tau_L} f(r) \qquad k \in {\mathbb Z}
\ee
where
\be
\label{LeftQuasi}
\tau_L = 1/4 \pi T_R(h_R + n) \qquad n = 0,1,2,\ldots
\ee
These solutions describe left-moving excitations which decay on a timescale
$\tau_L$.

We would like to emphasize that these quasinormal modes are
supergravity solutions which govern the late-time behavior of certain
bulk closed-string fields \cite{QuasinormalReview}.  They should not
be confused with the quasiparticles, which we regard as arising from
fundamental degrees of freedom on the stretched horizon.  But
we expect the quasiparticles to provide a holographic description of
low-energy excitations near the black hole, so their lifetimes should
govern the decay of these low-lying supergravity excitations.  It is
therefore very encouraging that the quasinormal lifetimes
(\ref{RightQuasi}),(\ref{LeftQuasi}) match our proposed quasiparticle
lifetimes.

We can gain more insight into the meaning of these quasinormal
frequencies by studying the dual CFT.  In a two-dimensional CFT the
finite-temperature Euclidean 2-point function of primary fields is
determined by conformal invariance:
\be
\label{CFTcorrelator}
\langle{\cal O}(w_1,\bar{w}_1) {\cal O}(w_2,\bar{w_2})\rangle \sim
\left(\pi T_L \over \sinh \pi T_L(w_1 - w_2)\right)^{2h_L}
\left(\pi T_R \over \sinh \pi T_R(\bar{w}_1 - \bar{w}_2)\right)^{2h_R}
\ee
where
\[
-\infty < {\rm Re} \, w < \infty \qquad {\rm Im} \, w \approx {\rm Im} \, w + 1/T_L \qquad
{\rm Im} \, \bar{w} \approx {\rm Im} \, \bar{w} + 1/T_R\,.
\]
If one looks at a slice of fixed Euclidean time, say ${\rm Im} \, w =
0$, then the 2-point function falls off exponentially with spatial
separation.  Thus at finite temperature there is a finite correlation
length \cite{CFTbook}.  Moreover the correlation function factorizes
into holomorphic (left-moving) and anti-holomorphic (right-moving)
pieces.  The holomorphic part of the correlation function shows that
the correlation length in the $x^+ = t + x$ light-front direction is
given by
\[
\xi^+ = 1 / 2 \pi T_L h_L
\]
while the anti-holomorphic part of the correlator shows that the
correlation length in the $x^- = t - x$ direction is
\[
\xi^- = 1 / 2 \pi T_R h_R\,.
\]
To measure a correlation length in the $x^+$ direction one needs to
consider excitations which propagate in $x^+$.  Thus the light-front
correlation length for right-moving excitations is governed by $T_L$,
while for left-moving excitations it is governed by $T_R$.  This is
the CFT origin of the somewhat puzzling interchange of left and right
in the quasiparticle lifetimes.

These correlation lengths can be seen in a more precise way from the results of
\cite{BSS2,Gubser}, who computed the Fourier transform of
the retarded Green's function in the CFT and found a set of poles at
\be
\label{RightCFT}
k_+ = {1 \over 2} (\omega - k) = - i 2 \pi T_L (h_L + n) \qquad n = 0,1,2,\ldots
\ee
The $n=0$ pole determines the correlation length in the $x^+$
direction, and therefore sets the lifetime of the right-moving
quasiparticles.  The poles with $n > 0$ govern the subleading
exponential fall-offs in the correlator (\ref{CFTcorrelator}).
Likewise there are poles at
\be
\label{LeftCFT}
k_- = {1 \over 2} (\omega + k) = - i 2 \pi T_R (h_R + n) \qquad n = 0,1,2,\ldots
\ee
which set the lifetime of the left-moving quasiparticles.

Let us make a few comments on these CFT results.  At finite
temperature even free field theories have finite correlation lengths.
So our discussion is perfectly applicable to any operator with
non-trivial scaling dimension, even in a free CFT.  Note that it is
important to distinguish between the correlation lengths and
thermalization times of the system.  With generic interactions one
would expect these scales to be about the same.  However, as we
mentioned, correlation lengths are finite even in free field theories
which do not thermalize.

These CFT results have a limited range of validity, since the
correlator (\ref{CFTcorrelator}) is only exact when the spatial
coordinate of the CFT is non-compact.  However we are interested in a
CFT which lives on a circle, since the $\phi$ coordinate of the BTZ
metric is periodic with period $2\pi$.  One can only trust the general
results (\ref{RightCFT}),(\ref{LeftCFT}) as long as $T_R,\,T_L \gg 1$.
Fortunately this is the regime where the BTZ black hole (as opposed to
thermal AdS) dominates the partition function
\cite{MaldacenaStrominger}.  For further discussion of this issue see
\cite{BSS2}.

\subsection{Additional comments}
\label{AdditionalComments}

It is interesting to compare the quasiparticle approach with that of
\cite{mastrom} where a low-energy effective string description of the
dynamics of four and five-dimensional black holes was deduced from
supergravity arguments. This picture was far more detailed than the
one we have presented.  For example in four and five dimensions
\cite{mastrom} argued for $(0,4)$ supersymmetry of the CFT. The
resulting CFT required interactions of the left and right moving
degrees of freedom to describe Hawking radiation and scattering from
the black hole.  Our quasiparticle picture is rather more crude, and
unlikely to reproduce the detailed low energy scattering predictions
of that model. On the other hand the quasiparticle picture is much
simpler in that everything is expressed in terms of quasi-free
particles, that is, particles with finite energies and lifetimes but
with no other interactions.  The quasiparticle picture suffices to
describe a limited class of dynamic observables, but as we emphasize
in this paper, it is applicable to a much wider range black holes and
it also makes the entropy -- area relationship manifest.

We would also like to make some comments on the extremal limit $r_+
\rightarrow r_-$.  For the BTZ geometry (\ref{BTZmetric}) the horizon
area remains non-zero in the extremal limit.  That is, for BTZ black
holes we do not face the difficulties with degenerate horizons that
will be discussed in section \ref{sensection}.  We therefore expect
that the basic quasiparticle description remains valid at extremality.
However working out a quasiparticle picture of the extremal limit
could be somewhat involved.  By staying away from extremality we kept
the occupation numbers $N_R,\,N_L \gg 1$, which enabled us to give a
simple thermodynamic treatment of the quasiparticle gas.  It also
allowed us to avoid the breakdown of the semiclassical approximation
at extremality discussed in \cite{PSSTW}.

Finally, the BTZ case makes it clear that the real part of the lowest
quasinormal mode frequency $k$ in (\ref{RightQuasiMode}),
(\ref{LeftQuasiMode}) cannot be simply identified with the average
quasiparticle energy.\footnote{This was not an issue for the
Schwarzschild black holes and the black D$p$-branes considered
previously, where the real part of the quasinormal mode frequency was
of order the quasiparticle energy.}  Following \cite{PSSTW}, the
energy gap above the extremal BTZ state is the energy associated with
the lowest nontrivial quasiparticle mode when $N_L \sim 1$, which
implies $T_L \sim 1/\ell$, so that $\Delta E_{\rm CFT} \sim 1/\ell$ or
$\Delta E \sim 1/\ell^2$.\footnote{Keeping in mind that we are working
in units where $8 G=1$, this matches the energy gap of
\cite{MalSusskind}.}  On the other hand, the real part of the
frequency of the lowest nontrivial quasinormal mode is $k=1$, which
suggests an energy gap $\Delta E_{\rm CFT} =1$. This is related to the
observation of \cite{MalSusskind} that multiply wound effective
strings are needed to correctly describe the low-energy degrees of
freedom.  The quasiparticles would then be periodic under $\phi
\rightarrow \phi + 2 \pi \ell$. The operators in the CFT that couple
to quasinormal modes are periodic under $\phi \rightarrow \phi + 2
\pi$, so must be built out of composites of the elementary
quasiparticles.  Though the quasiparticle energy gap $1/\ell \ll T_L,
T_R$ when $N_L,N_R \gg 1$, the average quasiparticle energies $T_L$,
$T_R$ can be deduced by assuming a thermal distribution of left and
right moving particles at temperatures $T_L$ and $T_R$ respectively
and using the equipartition theorem.

\section{Comments on degenerate horizons}
\label{sensection}

The quasiparticle picture we have developed clearly breaks down in the
limit of vanishing horizon area.  In this section we comment on some
of the subtleties which arise in developing a stretched horizon or
quasiparticle description of black holes with degenerate horizons.
The main observation is that for degenerate horizons the stretched
horizon no longer needs to be placed at the Planck temperature.  That
is, the arguments to that effect given in section 2 break down in the
degenerate limit.

A concrete example of a degenerate horizon is provided by Sen's black
holes which are the most general charged asymptotically flat black
hole solutions in heterotic string theory compactified on $T^6$
\cite{Sen:1994eb}.  We concentrate on electrically charged but
non-rotating solutions in the extremal or BPS limit.\footnote{Away
from extremality we expect a well-defined quasiparticle picture, with
multiple species of quasiparticles to account for the electric
charge.}  This example gives some lessons on the limits of validity of
the quasiparticle picture.

Sen pointed out that even though the classical horizon area of these
black holes vanishes, the microscopic entropy obtained by counting the
corresponding heterotic string states is non-zero \cite{Sen:1995in}.
To resolve this apparent contradiction he introduced an entropy
associated with the horizon area of the {\it stringy} stretched
horizon\footnote{The {\it stringy} stretched horizon is located at the
radius where the local temperature coincides with the Hagedorn
temperature.  See the appendix of \cite{Sen:1995in}.}, and showed that
this entropy coincides with the entropy of an elementary heterotic
string having the same charge and mass as the black hole.  Stretched
horizons for extremal black holes have been further considered in
\cite{LuninMathur}.

Sen's analysis seems to conflict with the present paper because we
need to put the stretched horizon at the {\it Planck} temperature in
order to derive the correct entropy--area relation.  The conflict can
be resolved by noting that the approximations used here and in
\cite{quasi_I} break down when the horizon area vanishes.  In
particular the Stefan-Boltzmann law is a thermodynamic formula which
doesn't apply when the size of the black hole is of order string
scale.  For Sen's extremal black holes the classical horizon area
vanishes and $\alpha'$ corrections to the spacetime action are
expected to be important in understanding radiation from the stretched
horizon.

This breakdown of the Stefan-Boltzmann law due to stringy corrections
is quite general: whenever the horizon area becomes of order string
scale the simple quasiparticle picture of the horizon breaks down.
Fortunately we have other tools for studying such small black holes.
In particular the Stefan-Boltzmann law breaks down at the Horowitz -
Polchinski correspondence point \cite{HorowitzPolchinski}, where one
can match to the spectrum of elementary string excitations.  Also in
the extremal limit one can use supersymmetric non-renormalization
theorems to study a microscopic description of the black hole.

\section{de Sitter space}

In this section we turn from black hole horizons to cosmological
horizons, and present a quasiparticle description of the static patch
of empty de Sitter space.

There has been considerable debate over the correct quantum
description of de Sitter space \cite{deSitterReview}.  Some groups
have argued that the inflationary patch of de Sitter is dual to a
conformal field theory, in which case the thermodynamic properties of
the static patch might arise along the lines of \cite{KabatLifschytz}.
Other groups have argued that de Sitter space is unstable
\cite{unstable}, in which case there could be no precise dual
description of the static patch.  But even if de Sitter space is
ultimately unstable the quasiparticle picture could be valid, in the
sense that it provides an effective description of the static patch
accurate on timescales short compared to the de Sitter lifetime.  This
would be analogous to the quasiparticle description of a Schwarzschild
black hole developed in section \ref{SchwarzschildSection}.

\subsection{de Sitter thermodynamics}

The Euclidean metric for $d$-dimensional de Sitter space is
\be
\label{deSittermetric}
ds^2 = \left(1 - {r^2 \over \ell^2}\right) dt_E^2
+ \left(1 - {r^2 \over \ell^2}\right)^{-1} dr^2 + r^2 d\Omega_{d-2}^2
\ee
where $\ell$ is the de Sitter radius, related to the cosmological
constant by $\Lambda = (d-2)(d-1)/2\ell^2$.  This space is smooth at
the horizon $r = \ell$ provided the Euclidean time coordinate is
periodically identified, $t_E \approx t_E + 2 \pi \ell$, with a
periodicity corresponding to the de Sitter temperature $T = 1/2 \pi
\ell$.

To compute thermodynamic quantities we follow \cite{SusskindUglum} and
evaluate the Euclidean partition function off-shell.  That is we
periodically identify $t_E \approx t_E + \beta$ but keep $\beta$
arbitrary; in general this leaves a conical singularity at the
horizon.  In the semiclassical approximation the partition function is
$Z = e^{-I}$ where the Einstein-Hilbert action
\[
I = - {1 \over 16 \pi G} \int d^dx \, \sqrt{g} (R - 2 \Lambda)\,.
\]
We will regard the action as consisting of two contributions.  There is a bulk
contribution
\bea
\nonumber
I_{\rm bulk} & = & - {1 \over 16 \pi G} \cdot {\beta \ell A \over d - 1} \cdot {2 (d-1) \over \ell^2} \\
\label{BulkAction}
& = & - {A \over 4 G} \cdot {\beta \over 2 \pi \ell}\,.
\eea
In deriving this we wrote the volume of Euclidean de Sitter space
\[
\int d^dx \, \sqrt{g} = {\beta \ell A \over d-1}
\]
where $A = 2 \pi^{(d-1)/2} \ell^{d-2} / \Gamma\big((d-1)/2\big)$ is
the area of the event horizon.  We also used the expression for the
bulk scalar curvature $R = 2 d \Lambda / (d - 2)$, with $\Lambda =
(d-2)(d-1)/2\ell^2$.  The action also gets a contribution from the
curvature singularity at the horizon.  The horizon contribution has
the universal form \cite{SusskindUglum}
\[
I_{\rm horizon} = - {A \over 4 G} \left(1 - {\beta \over 2 \pi \ell}\right)
\]
Corresponding to this decomposition of the action one finds two contributions to the
semiclassical energy and entropy of de Sitter space.  The bulk contributions are
\bea
\label{BulkThermo}
E_{\rm bulk} & = & \left.{\partial \over \partial \beta}\right\vert_{\beta = 2 \pi \ell} I_{\rm bulk} = - {A \over 8 \pi G \ell} \\
\nonumber
S_{\rm bulk} & = & \left.\left(\beta{\partial \over \partial \beta} - 1\right)\right\vert_{\beta = 2 \pi \ell} I_{\rm bulk} = 0\,.
\eea
The contributions from the horizon are
\bea
\label{HorizonThermo}
E_{\rm horizon} & = & \left.{\partial \over \partial \beta}\right\vert_{\beta = 2 \pi \ell} I_{\rm horizon} = {A \over 8 \pi G \ell} \\
\nonumber
S_{\rm horizon} & = & \left.\left(\beta{\partial \over \partial \beta} - 1\right)\right\vert_{\beta = 2 \pi \ell} I_{\rm horizon} = {A \over 4 G}\,.
\eea
Note that the total energy $E_{\rm bulk} + E_{\rm horizon}$ vanishes
\cite{GibbonsHawking}, while the entropy arises solely from the
horizon.

To understand this better we turn to the work of Susskind
\cite{Susskind}, who pointed out that from the string theory point of
view the bulk contribution arises at closed-string tree level, from
spherical worldsheets which do not touch the horizon.  We do not
expect the quasiparticles to capture the corresponding bulk energy.
The contribution from the singularity, on the other hand, arises from
spherical worldsheets which are pierced by the horizon.  In a
Hamiltonian framework such diagrams would describe open strings stuck
on the horizon \cite{Susskind}.  We take the quasiparticles to
correspond to these open string degrees of freedom, and therefore
expect the quasiparticle gas to reproduce the energy and entropy
associated with the horizon.

As usual, we introduce a stretched horizon at a radius where the
proper temperature is equal to the Planck temperature.  The proper
rate at which the stretched horizon emits energy in outgoing Hawking
radiation is then
\[
{dE_{\rm proper} \over dt_{\rm proper}} \sim A T_{\rm proper}^d\,.
\]
To work in terms of energy $E$ conjugate to the de Sitter time
coordinate $t$ we multiply by the appropriate redshift factors, and
find that the rate at which energy is radiated is given by
\be
\label{deSitterRate}
{dE \over dt} \sim {A \over \ell_{\rm Planck}^{d-2}} T^2\,.
\ee

\subsection{Quasiparticle description}

It is straightforward to give a quasiparticle description of the
static patch of de Sitter space.  We postulate that the number of
quasiparticles is given by the de Sitter entropy, $N \approx S_{\rm
horizon}$.  The quasiparticle gas has a temperature equal to the de
Sitter temperature $T = 1/2 \pi \ell$.  Each quasiparticle has an
energy $\epsilon \approx T$ and a lifetime $\tau \approx 1/T$.

It is trivial to see that the thermodynamics of such a quasiparticle
gas reproduces the energy and entropy (\ref{HorizonThermo}) associated
with the de Sitter horizon.  The quasiparticle description also
predicts the rate at which the horizon radiates energy.
\[
{dE \over dt} = N \epsilon / \tau
\]
This agrees with the Stefan-Boltzmann law (\ref{deSitterRate}),
provided one identifies the entropy of the quasiparticle gas $S \sim
N$ with the area of the horizon in Planck units.

\subsection{de Sitter quasinormal frequencies}

To provide further evidence for the proposed quasiparticle lifetimes
we proceed to study quasinormal modes in de Sitter space.  For
simplicity we consider a massless, minimally-coupled scalar field.

The wave equation reads
\[
\left(- \Bigl(1 - {r^2 \over \ell^2}\Bigr)^{-1} \partial_t^2
+ {1 \over r^{d-2}} \partial_r r^{d-2}\Bigl(1 - {r^2 \over \ell^2}\Bigr)\partial_r
+ {1 \over r^2} \nabla^2_{S^{d-2}}\right) \phi = 0\,.
\]
Changing to a dimensionless radial coordinate $\rho = r / \ell$, $0
\leq \rho \leq 1$ and separating variables $\phi(t,\rho,\Omega) =
e^{-i \omega t} \phi(\rho) Y_j(\Omega)$, where $Y_j$ is a spherical
harmonic on $S^{d-2}$ with angular momentum $j$, one obtains the
radial wave equation
\be
\label{wave}
\left({1 \over \rho^{d-2}} \partial_\rho \rho^{d-2} (1 - \rho^2) \partial_\rho
+ {\ell^2 \omega^2 \over 1 - \rho^2} - {j(j+d-3) \over \rho^2}\right)\phi(\rho) = 0\,.
\ee

Quasinormal frequencies are determined by requiring that $\phi$ be
smooth at $\rho = 0$ (the center of the static patch) and purely
outgoing at $\rho = 1$ (that is, exiting the static patch at the
horizon).  From (\ref{wave}) it's clear that the quasinormal
frequencies are proportional to the de Sitter temperature.
\[
\omega \sim 1/\ell \sim T
\]
The quasinormal frequencies were determined exactly in
\cite{deSitterQuasi}.  For $d \geq 3$, the unique solution which obeys
the correct boundary conditions is\footnote{Two dimensional de Sitter
space is conformal to a finite cylinder.  One can have left- and
right-moving waves on the cylinder, but there doesn't appear to be any
sensible definition of quasinormal modes.}
\beas
&& \phi(\rho) = \rho^j (1 - \rho^2)^{i \ell \omega / 2} F(\alpha,\beta,\gamma;\rho^2) \\
\noalign{\vskip 2mm}
&& \alpha = {j + d - 1 + i \ell \omega \over 2} \quad \beta = {j + i \ell \omega \over 2}
\quad \gamma = j + {d - 1 \over 2}
\eeas
where the quasinormal frequencies are given by
\[
\omega = - {i \over \ell} \, (j + 2n) \quad {\rm or} \quad
\omega = - {i \over \ell} \, (j + d + 2n - 1) \qquad
n = 0,1,2,\ldots
\]
(the second possibility is redundant if $d$ is odd).  It is curious
that the de Sitter quasinormal frequencies are purely imaginary.  We
do not have an intuitive understanding of this fact.

To summarize, quasinormal excitations in de Sitter space decay on a
timescale set by the inverse temperature, $\omega^{-1} \sim \ell \sim
1/T$, in agreement with our proposed quasiparticle lifetime.

\section{Conclusions}

Long ago it was argued that the simplest way to think about black
holes is in terms of a stretched horizon located some distance outside
the true event horizon \cite{MembraneParadigm}.  The stretched horizon
has been shown to be an equally useful concept in the quantum theory
\cite{stu}.  In this paper we have analyzed several examples of black
hole and cosmological horizons, and shown that many properties of the
stretched horizon can be understood in terms of a simple gas of
non-interacting quasiparticles.  The quasiparticle gas reproduces the
equilibrium thermodynamics of the horizon.  In particular it naturally
accounts for the universal Bekenstein-Hawking relation between entropy
and horizon area.  Moreover, by relating the lifetime of the
quasiparticles to the imaginary part of the lowest quasinormal
frequency, we have shown that the quasiparticle gas correctly
describes thermalization on the horizon.  In particular it correctly
accounts for the universal thermalization relations
(\ref{Universal1}), (\ref{Universal2}).

There are several interesting directions in which the quasiparticle
picture could be extended.  It would be worthwhile to work out the
detailed quasiparticle description of an extremal BTZ black hole.  It
would also be interesting to study charged black holes, such as the
Reissner-Nordstrom solution or the Sen black holes mentioned in
section \ref{sensection}.  Presumably charged black holes are
analogous to the rotating BTZ solution, in that more than one species
of quasiparticle is required.  Finally, it would be interesting to
develop a quasiparticle picture for Rindler space.  This would
underscore the universality of the quasiparticle description, since
all macroscopic horizons are locally approximately Rindler.  In
attempting this one faces a puzzle, that the redshift factor at the
Planck stretched horizon of a Schwarzschild black hole
\[
1 / \sqrt{-g_{tt}} = T_{\rm Planck} / T \sim (\ell_{\rm Planck} M)^{1/(d-3)}
\]
diverges in the Rindler limit $M \rightarrow \infty$.  A natural way
to avoid this difficulty would be to work in terms of proper
quantities measured at the stretched horizon.  Indeed one might
ultimately hope to derive the quasiparticle description starting from
an action principle for a membrane located at the stretched horizon
\cite{Maulik}.

\bigskip
\centerline{\bf Acknowledgements}
\noindent
We are grateful to Gary Horowitz, Samir Mathur and Lenny Susskind for
valuable discussions.  NI and DK are supported by the DOE under
contract DE-FG02-92ER40699.  The research of DL is supported in part
by DOE grant DE-FE0291ER40688-Task A.  NI, DK, DL and GL are supported
in part by US--Israel Bi-national Science Foundation grant \#2000359.
D.L. thanks the Departamento de F\'{i}sica, CINVESTAV for hospitality
during the completion of this work.

\appendix
\section{Numerical algorithms for quasinormal frequencies}

We use Mathematica to solve for quasinormal mode frequencies. The wave
equation (\ref{numer}) is solved using a power series expansion about
the regular singular point $z=0$. The coefficients are determined
using a recursion relation, as shown in the code below, and the
boundary condition (\ref{horbc}) is selected. This solution is matched
at a regular point (e.g. $z=1/2$), with a numerical solution of
(\ref{numer}) obtained via the Runge-Kutta method, starting from
$z=1-\epsilon$, $\epsilon \ll 1$, with the boundary condition
(\ref{rooteq}). The matching involves setting $f'(z)/f(z)$ equal,
which is achieved by iteratively adjusting $\rho$ using Mathematica's
FindRoot routine.  The FindRoot routine uses an initial guess for the
frequency coming from solving (\ref{rooteq}) using the power series
method \cite{hubeny} truncated to 10 terms.

\begin{verbatim}
b1[0]=0.5581972109-0.9281852713 I;
b1[1]=0.6221606713-0.8830128157 I;
b1[2]=0.6985155813-0.8030747084 I;
b1[3]=0.7673368820-0.6578525677 I;
b1[4]=0.7647605329-0.4398341076 I;

n=400; p=0; xx = 1/2;  (* Adjustable parameters: n= number of
terms in power series, p = p-brane, xx = matching point *)

(* Construct power series solution using recursion relation *)
ser=Series[1/(1-x)^((9-p)/(7-p)),{x,0,n}];
Do[b[l]=SeriesCoefficient[ser,l],{l,0,n}];
f[rho_]:=Module[{},a[0]=1;
          Do[a[k]=N[-1/((k-I rho)^2+rho^2) Sum[a[k-l]b[l]rho^2,
           {l,1,k}],100] ,{k,1,n}];
          N[Sum[(k-I rho) a[k] (xx)^(k-I rho -1),{k,0,n}],100]/
          N[Sum[a[k] (xx)^(k-I rho),{k,0,n}],100]];


(* Solve ODE using Runge-Kutta *)
eps=10^(-25);
g[rho_] := Module[ {},
            ans = NDSolve[ {w^((9-p)/(7-p)) ((1-w)^2 D[chi[w],w] -
             (1-w)chi[w]) + rho^2 phi[w] == 0, D[phi[w],w]==chi[w],
             phi[eps]==0,chi[eps]==1}, {phi,chi}, {w,3/5,eps},,
             WorkingPrecision->30,Method->RungeKutta,MaxSteps->10000];
            N[-chi[1-xx]/phi[1-xx] /. ans][[1]]];

FindRoot[ g[rho]-f[rho]==0, {rho,b1[p],b1[p]-.01}, AccuracyGoal->12,
  MaxIterations->15]   (* Output is quasinormal mode frequency rho *)

\end{verbatim}

\section{General $p$-branes}

\subsection{Decoupling limits}
\label{GeneralDecoupling}

We start with black p-brane solutions of the gravity action in $D$
dimensions \cite{CveticYoum,lupope}
\be
S=-\frac{1}{2\kappa^{2}}\int d^{D}x \sqrt{g}[R-\frac{1}{2}(\partial
\phi)^{2}
-\frac{1}{2(d+1)!}e^{a\phi}F^{2}_{d+1}]
\ee
where $D=p+d+3$.  The solution with one harmonic function takes the form (in string frame)
\be
ds^2=H^{\alpha-\frac{aN}{4}}(r)(H^{-N}(r)[-f(r)dt^2+dy_{1}^2 + \cdots dy_{p}^2]
+f^{-1}(r)dr^2 +r^2 d\Omega_{d+1}^2)
\ee
where
\be
H(r)=1+\frac{\mu^{d}\sinh^{2}\gamma}{r^d} \qquad
f(r)=1-\frac{\mu^{d}}{r^{d}}\,.
\ee
The dilaton and gauge potential are
\bea
&&e^{2(\phi(r)-\phi(\infty))}=H^{-aN}\\
\nonumber
&&F_{d+1}=\frac{1}{2}\sqrt{N}d\mu^{d}\sinh 2\gamma \epsilon_{d+1}
\eea
and the parameters have a relationship
\be
\alpha=\frac{N(p+1)}{D-2}, \quad N=4[a^2+\frac{2d(p+1)}{D-2}]^{-1}\,.
\ee
In the near extremal limit the charge and thermodynamical
functions (per unit volume) are given by \cite{kt}
\bea
\nonumber
q_{p} &= & \frac{\omega_{d+1}}{2\sqrt{2}\kappa}d\sqrt{N}\mu^{d}\sinh 2
\gamma\\
E &=& \frac{d\lambda\omega_{d+1}}{\sqrt{N}\kappa^{2}}\mu^{d}\\
\nonumber
S &= & 4\pi
d^{-(d+1)/d}\lambda^{-\lambda}\omega_{d+1}^{-1/d}(\sqrt{2}\kappa)^{2/d
-N/2}(\frac{q_{p}}{\sqrt{N}})^{N/2} E^{\lambda}
\eea
where
\be
\lambda=\frac{d+1}{d}-\frac{N}{2}\,.
\ee
We want to go to a near horizon limit for which we can define a
decoupled theory. Let us introduce a length scale that we will call
$l_s$ (which would be taken to zero at the end), this could be the
string length but could also be the Planck length.

We will change variables 
\be
r=U^{2b-1}l_{s}^{2b}
\ee
where $b$ at the moment is a free parameter.  Another free parameter
(at the moment) is $\beta$ where
\be
\mu^{d}e^{2\gamma}=g^{2} l_{s}^{2\beta}
\ee
where $g^2$ is some dimensional parameter which is kept fixed in the
limit $l_{s} \rightarrow 0$.  To get a finite relationship between
energy and horizon size one has to fix in this limit a parameter which
we call $g_{ym}$, which satisfies
\be
\kappa l_{s}^{-bd}=g_{ym}^{2}.
\ee
To get a metric which scales like $l_{s}^{2}$ (in the string frame)
one has two conditions
\bea
&&N(\beta-db)=-2b \nonumber\\
&&\frac{a}{4}=\frac{p+1}{D-2}-1+\frac{1}{2b}
\label{decmet}
\eea
We also want to  get a finite dilaton $e^{\phi}$ after decoupling.
For this we need to know how $g_{s}=e^{\phi(\infty)}$ scales in the
decoupling limit. In $D$ dimensions
\be
\kappa \sim \frac{g_{s} l_{s}^{4}}{\sqrt{V_{10-D}}}
\ee
where $V_{10-D}$ is the volume of the $10-D$ dimensions which one has
compactified on (not to be confused with $y_{1} \cdots y_{p}$).
If we do not wish to introduce any more dimensionful parameters then
we should take
\be
V_{10-D} \sim l_{s}^{10-D}\,.
\ee
In this case the condition for a finite dilaton is  
\be
\frac{a}{4}=\frac{1}{2b}-\frac{d}{8}+\frac{D-10}{16b}\,.
\label{decdil}
\ee
Equations (\ref{decmet}), (\ref{decdil}) have solutions only if $D=10$
for specific values of $a$, or in $D \neq 10$ with $a=0$.  This means
that in the decoupling limit apart from the ten dimensional branes the
only other decoupled solutions are those with a constant dilaton.
Indeed if there is no dilaton ($a=0$, or M-branes) then the equations
always have a solution
\be
b=\frac{D-2}{2(D-p-3)}\,.
\ee
It turns out that in all cases the conditions (\ref{decmet}),
(\ref{decdil}) are enough to insure a finite relationship between
energy density and entropy density.

The metric that one gets after this procedure (in Einstein frame) is
given by
\bea
&& ds^{2} = l_{s}^{2} (\frac{g^2}{U^{d(2b-1)}})^{\alpha-N}[-(1-(\frac{U_{h}}{U})^{d(2b-1)})dt^2 +
dy_{i}^{2}\label{newmet} \\
&&\quad + (1-(\frac{U_{h}}{U})^{d(2b-1)})^{-1}g^{2N}U^{(2b-1)(2-Nd)-2}dU^2
+g^{2N}U^{(2b-1)(2-Nd)}d\Omega_{d+1}^2]
\nonumber
\eea
We also have the temperature -- energy relationship
\be
T^{-1} \sim e^{N} g_{ym}^{2(2/d-N)}E^{\lambda-1}
\ee
and the energy -- radius relationship
\be
E \sim g_{ym}^{-4} U_{h}^{d(2b-1)}
\ee
which gives a temperature -- radius relationship
\be
T^{-1} \sim g^{N}U_{h}^{d(2b-1)}
\ee
that is independent of what we called $g_{ym}$.

We now define a new coordinate $\rho$
\be
\rho = g^{N}U^{d(2b-1)(\lambda-1)}
\ee
and the Einstein metric takes the form
\bea
ds^2 &=& l_{s}^{2}(\frac{g^{N/(\lambda-1)+1}}{\rho^{1/(\lambda-1)}})^{\alpha-N}
[-(1-(\frac{\rho_{h}}{\rho})^{1/(\lambda -1)})dt^{2}
+ dy_{i}^{2} \nonumber \\
&+& \frac{2d^{2}(2b-1)}{2-dN}(1-(\frac{\rho_{h}}{\rho})^
{1/(\lambda -1)})^{-1} d\rho^2 + \rho^{2} d\Omega_{d+1}^2]
\label{newmet1}
\eea

\subsection{Quasinormal modes}

After changing variables to $z=1-(\frac{\rho_{h}}{\rho})^
{1/(\lambda -1)}$, the wave equation for a minimally coupled scalar in
the metric (\ref{newmet1}) becomes
\be
\label{GeneralWave}
z \frac{\partial}{\partial z} \left( z \frac{\partial f}{\partial z}
\right) + \rho^2 (1-z)^{-2 +N +
\frac{2}{3-D+p} } f(z) = 0
\ee
where
\be
\rho^2 = \frac{\omega^2 e^{2N} \mu^{2-Nd} l_s^\delta}{d^2} 
\ee
and $\delta$ may be read off from (\ref{newmet1}). Again we find
that $\omega$ is proportional to the Hawking temperature of the black
brane.  Let us now consider supersymmetric examples which are
non-dilatonic ($a=0$) for $D<10$, generalizing the results of section
\ref{bbranes}.

For $D=6$, $p=1$, $N=2$ the wave equation is soluble analytically. The
solution ingoing on the future horizon is
\be
f(z) = z^{-i \rho} {}_2 F_1( -i \rho, -i \rho, 1-2 i \rho, z)\,.
\ee
Solving the Dirichlet boundary condition at infinity leads to the
equation
\be
1-i\rho = -j
\ee
where $j$ is a positive integer, so the quasinormal mode frequencies
are $\rho = -i (j+1)$.

For $D=5$, $n=3$, $p=0$ we get the same wave equation for the lowest
mode as when $D=4$, $n=4$, $p=0$. The decoupled geometry is simply
$AdS_2$ times a sphere of constant size. The wave equation is again soluble
analytically 
\be
f(z) = A e^{i \rho \log z} + B e^{-i  \rho \log z}
\ee
with $A$ and $B$ constants of integration. There are therefore no
quasinormal modes, only purely ingoing or outgoing solutions.

\subsection{Quasiparticle description}

In the above cases we find that the quasinormal mode frequency is
proportional to the Hawking temperature of the black brane. We can
then try to reproduce the thermodynamic properties of the black brane
using a quasiparticle picture.  We postulate that the microscopic
degrees of freedom giving a holographic description of the black hole
(or more generally the degrees of freedom that make up the stretched
horizon of the black brane) can be treated as a gas of quasi-free
particles in one lower dimension.  We associate a quasiparticle
lifetime with the imaginary part of the lowest quasinormal mode
frequency. We assume the number of quasiparticles accounts for the
entropy of the black brane, and that each quasiparticle carries energy
of order $T$.  The entropy is power-law in the temperature,
\be
\label{entrop}
S\propto E^\lambda\,.
\ee
It follows that the thermalization rate is
\be
\frac{dE}{dt} \propto A T^2~.
\ee
This agrees with a calculation of the same quantity using the
Stefan-Boltzmann law, assuming thermal radiation is emitted off a
stretched horizon one Planck distance from the event horizon, as in
\cite{quasi_I}.

The $D=5$, $n=3$, $p=0$ black hole and its cousin the $D=4$, $n=4$,
$p=0$ case are rather problematic in the decoupling limit.  In each
case the geometry is $AdS_2$ times a sphere. In these cases
$\lambda=0$, so (\ref{entrop}) implies the temperature diverges.  The
decoupling limit is rather degenerate in this case, as discussed in
\cite{frag} where more general decoupling limits are also considered.
Furthermore, quantum corrections violate decoupling of the $AdS_2$
throat and the asymptotically flat region \cite{sprad}. In these cases
the properties of the quasiparticles cannot be deduced from the
classical geometry alone.  However if we include the asymptotically
flat region, we expect to get well defined quasinormal modes, and we
should be able to construct a sensible quasiparticle description.  It
would be interesting to examine the quasinormal modes of the general
asymptotically flat $D=4,5$ black holes constructed in \cite{hms,hlm}
and develop a quasiparticle picture for these cases.  In these cases,
we expect that multiple species of quasiparticles are needed to
correctly describe the low energy dynamics of the black holes, as was
the case for the BTZ black hole that we analyzed in detail.


\end{document}